\documentclass[conference, a4paper]{IEEEtran}
\IEEEoverridecommandlockouts

\ifCLASSINFOpdf
  \usepackage[pdftex]{graphicx}
  \DeclareGraphicsExtensions{.pdf,.jpeg,.png}
\else
\fi

\usepackage{cite}
\usepackage{amsmath,amssymb,amsfonts}
\usepackage{algorithmic}
\usepackage{graphicx}
\usepackage{textcomp}
\usepackage[left=1.4cm, right=1.4cm, top=1.7cm]{geometry}
\usepackage{caption}
\usepackage{anyfontsize}
\usepackage{url}
\usepackage[table,xcdraw]{xcolor} 
\usepackage{tabularx} 
\usepackage{enumitem} 
\newcommand{\pquote}[2]{\textit{``#1''}{\,}{\small[#2]}}
\usepackage[os=mac]{menukeys}
\usepackage{xspace}

\setlist[itemize]{leftmargin=*}
\setlist[enumerate]{leftmargin=*,label=\arabic*.}

\newcommand{\sd}[1]{{\small\mbox{{\textsc{sd}}=#1\xspace}}}
\newcommand{\mean}[1]{{\small\mbox{{\textsc{m}}=#1\xspace}}}
\newcommand{\median}[1]{{\small\mbox{median}=#1\xspace}}

\newcommand{\chisquare}[5]{{\small$\chi^2(#1,N=#2)=#3$, $p#4#5$}}

\makeatletter
\newcommand*\titleheader[1]{\gdef\@titleheader{#1}}
\AtBeginDocument{%
\let\st@red@title\@title
\def\@title{%
\bgroup\normalfont\normalsize\centering\@titleheader\par\egroup
\vskip0.2em\st@red@title}
}
\makeatother

\makeatletter
\renewcommand{\fnum@figure}{Figure \thefigure}
\makeatother

\title{{Do ATCOs Need Explanations, and Why?} \\
\centering{\Large{Towards ATCO-Centered Explainable AI for Conflict Resolution Advisories}} 
\vspace{0.5cm}
}

\titleheader{First US-Europe Air Transportation Research and Development Symposium (ATRDS 2025)}

\author{
    \IEEEauthorblockN{Katherine Fennedy$^1$, Brian Hilburn$^3$, Thaivalappil N.M. Nadirsha$^1$, Sameer Alam$^{1,2}$, Khanh-Duy Le$^4$, Hua Li$^{1,2}$}\\ 
    \IEEEauthorblockA{
        \begin{minipage}[t]{0.4\textwidth}
            \begin{minipage}[t]{\textwidth}
                \centering
                \vspace{-1.5em} 
                \textsuperscript{1}Air Traffic Management Research Institute
            \end{minipage}
            \begin{minipage}[t]{\textwidth}
                \centering
                \vspace{-1.5em} 
                \textsuperscript{2}School of Mechanical \& Aerospace Engineering\\
                Nanyang Technological University, Singapore\\
                \{katherine.fennedy, mohdnadirsha.tn,\\
                sameeralam, lihua\}@ntu.edu.sg
            \end{minipage}
        \end{minipage}
        \hfill
        \begin{minipage}[t]{0.35\textwidth}
            \centering
            \vspace{-1.5em} 
            \textsuperscript{3}Center for Human\\
            Performance Research\\
            Philadelphia, United States\\
            brian.hilburn@chpr-usa.com
        \end{minipage}
        \hfill
        \begin{minipage}[t]{0.25\textwidth}
            \centering
            \vspace{-1.5em} 
            \textsuperscript{4}VNUHCM\\
            University of Science\\
            Ho Chi Minh City, Vietnam\\
            lkduy@fit.hcmus.edu.vn
        \end{minipage}
    }
}

\renewcommand\IEEEkeywordsname{Keywords}
\IEEEaftertitletext{\vspace{-1\baselineskip}}

\begin{document}

\maketitle

\noindent

\begin{abstract}

Interest in explainable artificial intelligence (XAI) is surging.
Prior research has primarily focused on systems’ ability to generate explanations, often guided by researchers’ intuitions rather than end-users' needs.
Unfortunately, such approaches have not yielded favorable outcomes when compared to a black-box baseline (i.e., no explanation). 
To address this gap, this paper advocates a human-centered approach that shifts focus to air traffic controllers (ATCOs) by asking a fundamental yet overlooked question: Do ATCOs need explanations, and if so, why?
Insights from air traffic management (ATM), human-computer interaction, and the social sciences were synthesized to provide a holistic understanding of XAI challenges and opportunities in ATM.
Evaluating 11 ATM operational goals revealed a clear need for explanations when ATCOs aim to document decisions and rationales for future reference or report generation.
Conversely, ATCOs are less likely to seek them when their conflict resolution approach align with the artificial intelligence (AI) advisory.
While this is a preliminary study, the findings are expected to inspire broader and deeper inquiries into the design of ATCO-centric XAI systems, paving the way for more effective human-AI interaction in ATM.

\end{abstract}

\vspace{0.3cm}


\section{Introduction}
Rapid integration of artificial intelligence (AI) into air traffic management (ATM) is transforming decision-making processes by enhancing both safety and efficiency.
However, the inherently safety-critical nature of ATM demands not only technical performance from AI systems but also robust explainability—a need underscored by regulatory bodies like the European Union Aviation Safety Agency (EASA).
EASA defines explainability as the ``capability to provide the human with understandable, reliable, and relevant information with the appropriate level of detail and with appropriate timing on how an AI application produces its results''\cite{easa2024}.
It distinguishes between two types of explainability: operational explainability for end users, and development explainability for system designers and auditors.
This work focuses on the former, addressing the needs of air traffic controllers (ATCOs) for explainable AI (XAI).

\begin{figure}[t]
    \centering
    \includegraphics[width=\columnwidth]{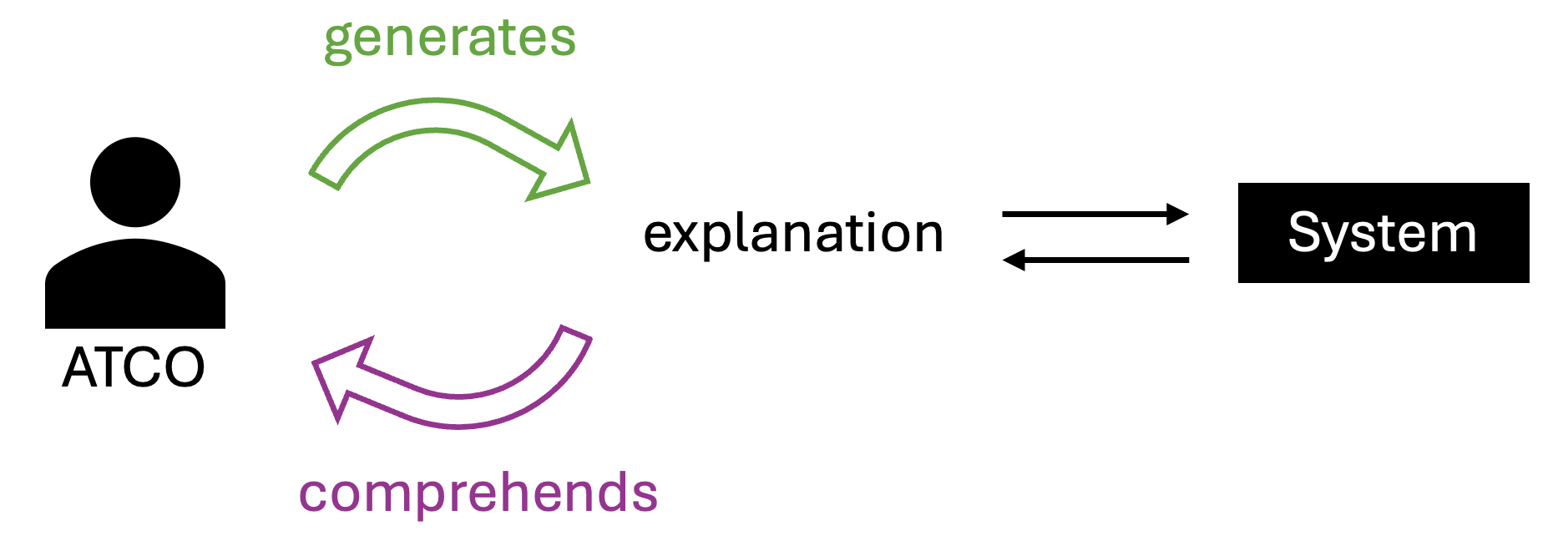}
    \caption{Explanation as an enabler in ATCO-AI interaction}
    \label{fig:atco-system}
\end{figure}

Integrating XAI into ATM is crucial for several key reasons.
Firstly, XAI has been recognized as one of the fundamental building blocks for a trustworthy AI\cite{easa2024}, an essential attribute in safety-critical domains like ATM.
Beyond merely improving user acceptance, AI integration in ATM is expected to enhance human decision-making performance by providing ATCOs with meaningful context-aware explanations.
Furthermore, if ATCO-generated explanation can be integrated into the AI system—clarifying why one advisory is accepted while another is rejected—the system stands to benefit by learning from its users (see Fig.~\ref{fig:atco-system}).
Overall, XAI is envisioned to serve a critical role in enabling a more effective human-AI interaction in ATM~\cite{kirwan2025human}, from trust calibration to supporting co-evolution~\cite{pham2024hah}.

Recent research in XAI for ATM has made significant strides by developing and validating prototypes that offer various visual explanations.
However, evidence indicates that system-generated explanations may fall short if they do not align with ATCOs’ operational needs~\cite{jmoona2023tot}.
Insights from human-computer interaction (HCI) frameworks further suggest that explanation design should begin with a clear understanding of users’ reasoning goals~\cite{jin2021practice}, rather than retrospectively validating a pre-designed interface.
Building on these perspectives, the proposed ATCO-centered study reverses the conventional top-down approach by starting with the fundamental question of why ATCOs need explanations.
This emphasis on user-generated insights grounded in their training and operational experience~\cite{wang2019theory}, allows us to capture the nuanced motivations behind their needs for explanations.

    \begin{table*}[t]
        \centering
        \renewcommand{\arraystretch}{1.5} 
        \setlength{\tabcolsep}{5pt} 
        \begin{tabularx}{\textwidth}{|p{1.7cm}|X|X|} 
        \hline
        \textbf{} & \textbf{\cellcolor{black}\color{white}Transparency} & \textbf{\cellcolor{black}\color{white}Explainability} \\ \hline
        
        \textbf{\cellcolor{black}\color{white}Responsibility} & 
        Revealing internal processes (algorithm, parameters) of the system &
        Ensuring the processes and outputs are understandable \\ \hline
        
        \textbf{\cellcolor{black}\color{white}Question} &
        \textbf{What} the system is doing? \textbf{How} it works? &
        \textbf{Why} the system did what it did? \textbf{Why not} something else? \\ \hline
        
        \textbf{\cellcolor{black}\color{white}Purpose} &
        Provide visibility into the system &
        Provide clarity and usability to the user \\ \hline
        
        \textbf{\cellcolor{black}\color{white}Relationship} &
        Transparency is the foundation, which explainability builds on &
        Explainability enhances transparency by making it accessible \\ \hline
        
        \end{tabularx}
        \caption{Highlighting the differences between the term \textit{transparency} and \textit{explainability}.}
        \label{tab:transparency_vs_explainability}
    \end{table*}

Guided by Jin et al's framework~\cite{jin2021practice}, we conducted user interviews, explored 11 goals, and employed goal-sorting exercises to identify key motivations for seeking explanations and to examine how explanations influence the acceptance or rejection of AI advisory.
For instance, all eight ATCO participants needed explanations when generating reports, while seven indicated that explanations are essential for communicating their decisions to supervisors.
Additionally, six ATCOs reported a strong need for explanations when the goal is to learn from AI or to differentiate between similar instances.
On the other hand, ATCOs were less likely to need explanations when their assessments aligned with the AI's advisory.
Importantly, the findings reveal the importance of dynamically adjusting explanations to promote appropriate trust calibration, ensuring that explanations effectively support human-AI collaboration in ATM.


\section{Related Work}

    \subsection{XAI Research Landscape in ATM}
    In a recent literature review, Degas et al.~\cite{degas2022survey} highlighted two insights that warrant our consideration.
    Firstly, they advocated that XAI for ATM should move towards a more user-centric design, in which AI and user can understand and interact with one other effectively.
    Current research has tended to focus on what AI systems can do, rather than the needs of the user~\cite{miller2019socialscience,hernandez2021framework}.
    For instance, explanations can be either global (e.g. full decision tree) or local (e.g. a single branch), and are often based solely on researchers' intuition.
    While designing and implementing XAI models are beyond current scope, the present work places ATCOs at the center of the research, co-designing solutions and deriving insights that reflect their experiences and operational needs.
    
    Secondly, Degas et al.~\cite{degas2022survey} echoed Arrieta et al.~\cite{arrieta2020xai} by highlighting the common conflation of the terms \textit{transparency} and \textit{explainability}, and its detrimental impact on XAI development.
    While the distinction between these terms may appear subtle, it is important to delineate them clearly to avoid perpetuating confusion in future research.
    Table~\ref{tab:transparency_vs_explainability} attempts to clarify their differences and similarities.
    For a deeper discussion with other often conflated terms like `comprehensibility' and `interpretability', readers are referred to~\cite{arrieta2020xai}.
    Thus, to maintain conceptual clarity, the terms \textit{transparency} and \textit{explainability} are enclosed in single quotation marks throughout this paper~\footnote{This paper focuses on general approaches to explainability (e.g. clarity, contextual examples, iterative communication) within the context of human-human interaction, and not specific XAI methods (e.g. SHAP, LIME), except where explicitly mentioned}, preserving the original intent of the cited authors.

    Transparency and explainability are often \textit{correlated} but not \textit{causally} linked.
    For instance, a conflict detection tool can be transparent without being explainable (e.g. revealing a neural network's algorithm too complex for ATCOs to understand), or explainable without being transparent (e.g., stating ``converging flight paths at 30,000 feet in five minutes'' without revealing internal logic due to proprietary reasons).
    These concepts can be seen as two sides of the same coin, both essential for building trust in AI systems.
    Transparency connects to the model's back-end, while explainability serves as the front-end engaging the user more actively.
    
    \subsection{XAI Visualizations in ATM}
    Over the past five years, research on ATM XAI visualizations has made notable strides, ranging from conceptual interface designs to functional prototypes validated empirically with ATCOs.
    One early concept, proposed by Xie et al.~\cite{xie2021xai}, explains the risk of incidents and accidents based on meteorological data using predictive models at both global and local scales.
    The interface was designed for a secondary monitor to avoid cluttering the primary tactical radar display.
    Pushparaj et al.~\cite{pushparaj2023xai} offered another perspective by demonstrating that the presence of explanations influenced brain activity and trust.
    Nevertheless, the study found no evidence of an impact on `understandability' of the provided explanations or on their ability to address ATCOs' specific questions—potentially confounding the findings with broader cognitive human factors reported in the paper.

    The recently completed SESAR-funded TAPAS (Towards an Automated and exPlainable ATM System) and ARTIMATION (Transparent AI and Automation To ATM Systems) projects both align closely with our research objectives.
    The TAPAS project developed and validated a prototype for each of two use cases: 1) Air Traffic Flow and Capacity Management and 2) Conflict Detection and Resolution (CDR), both of which differ significantly in their safety and time-critical requirements.
    TAPAS uncovered key insights into \textbf{when} and \textbf{how} `explanations' should be provided for systems to be acceptable and trustworthy for users~\cite{valle2022TAPAS}.
    However, the project also acknowledged that, there was limited motivation at that time to distinguish conceptually between `explainability' and `transparency' when defining functional requirements for the prototype~\cite{tapasD3_1}.

    Overlapping with the TAPAS project, ARTIMATION investigated how `transparent' algorithms could help ATCOs better understand and accept solutions in two prediction use cases: CDR and Take-Off Time (TOT) delay.
    While TAPAS focused on evaluating prototypes against a usability checklist, ARTIMATION focused on comparing distinct approaches to presenting `explanations'.
    
    In the CDR use case, ARTIMATION compared three visual representations: 1) Black Box (BB), 2) Heat Map (HM), and 3) Story Board (SB)~\cite{hurter2022cdr}.
    BB was the simplest as it displayed only the algorithm's proposed solution without explanation, serving as the baseline.
    HM built on BB by rendering green and red envelopes to indicate whether the algorithm's explored solution was good or bad.
    SB combined multiple layers of information, including a) a timeline of aircraft trajectory changes, b) a less efficient alternative solution, and c) suggested actions for delayed implementation.
    Interestingly, these approaches may better reflect varying levels of transparency rather than the claimed levels of `explanation', as they progressively reveal more information without guaranteeing improved user comprehension.

    ARTIMATION's second use case compared various XAI models (SHAP, LIME, DALEX) using breakdown plots to illustrate key features influencing TOT delay.
    Factors contributing to increased delay were shown as red bars, those reducing delay as green bars, and the final predicted delay as blue bar.
    Among the models, DALEX was judged to be more usable and was preferred by ATCOs.
    However, all three models received negative feedback for failing to convey the `operational relevance' of the selected features~\cite{jmoona2023tot}.
    Such feedback highlights that the effectiveness of explanations depends not only on \textbf{how} they are presented but also on \textbf{what} elements are included in the explanation.

    Another SESAR-funded project relevant to our work is MAHALO (Modern ATM via Human/Automation Learning Optimisation).
    It investigated the effects of both `conformance' and `transparency' on ATCOs' acceptance, agreement, workload, and subjective feedback~\cite{mahaloD6_2}.
    For `transparency', the project tested three conditions: vector serving as the baseline, diagram illustrating the solutions space, and text (which combined both solution space and a table detailing the `when', `what', `how', and `why' of automation activities).
    While varying levels of `conformance' produced main effects, `transparency' did not.
    Although several attempts have been made to provide system-generated explanations in ATM, most have assumed that ATCOs inherently need explanations.
    The present study aims to address the lack of evidence supporting the assumed need by investigating whether and why ATCOs need explanations.

    \subsection{Human-Centered XAI Insights from Non-ATM Domains}
    The relatively mature field of HCI offers valuable theoretical and practical frameworks that have the potential to be applied to ATM, given its strong user-centered focus.
    Two major works are particularly relevant to the current scope.
    
    First, Wang et al.~\cite{wang2019theory} proposed a conceptual framework that outlines pathways through which specific explanations can support reasoning, identifies how certain reasoning methods may fail due to cognitive biases, and suggests how XAI elements can mitigate these failures.
    The framework recommends first considering users’ reasoning goals and biases, informed through literature review or participatory design methods.
    Next, it advises identifying explanations that support these reasoning goals or address biases using the proposed pathways.
    Finally, it emphasizes integrating XAI capabilities into explainable user interfaces.
    This approach contrasts with many ATM studies, which often begin by designing and implementing interfaces based on researchers’ assumptions about what constitutes a ``good'' explanation~\cite{miller2019socialscience}, involving users only during the final validation phase—a stage that may be too late.

    Second, while Wang et al.~\cite{wang2019theory}’s framework takes a theory-driven approach, Jin et al.~\cite{jin2021practice} offered a complementary, practical approach focusing on study execution.
    Their framework—comprising explanation goals and an iterative prototyping workflow—serves as an expedient method to help XAI system designers understand user- and scenario-specific requirements. 
    To the best of our knowledge, the present study represents the first application of Jin et al.'s framework in the ATM domain.

\section{Proposed ATCO-Centered Approach}
Leveraging insights from related work, this study aims to adopt best practices from non-ATM fields to advance XAI development in the ATM domain.
The approach is distinguished by two key characteristics: 1) this study positions ATCOs and their operational expertise at the core of the investigation, ensuring their insights guide the process from the outset, and 2) this study reverses the conventional top-down approach found in the literature by starting with the fundamental question of `why.'

Both the TAPAS and ARTIMATION projects relied on prototype validation and system-generated explanations to gather insights from ATCOs.
In contrast, the current approach focuses on human-generated explanations as the foundation of our investigation.
This focus draws on insights from human-human interaction and applies them to the envisioned context of ATCO-AI interaction.
For instance, the study leverages natural processes already familiar to ATCOs by understanding the role of explanations they received during their training.

\begin{figure}[b]
    \centering
    \includegraphics[width=0.75\columnwidth]{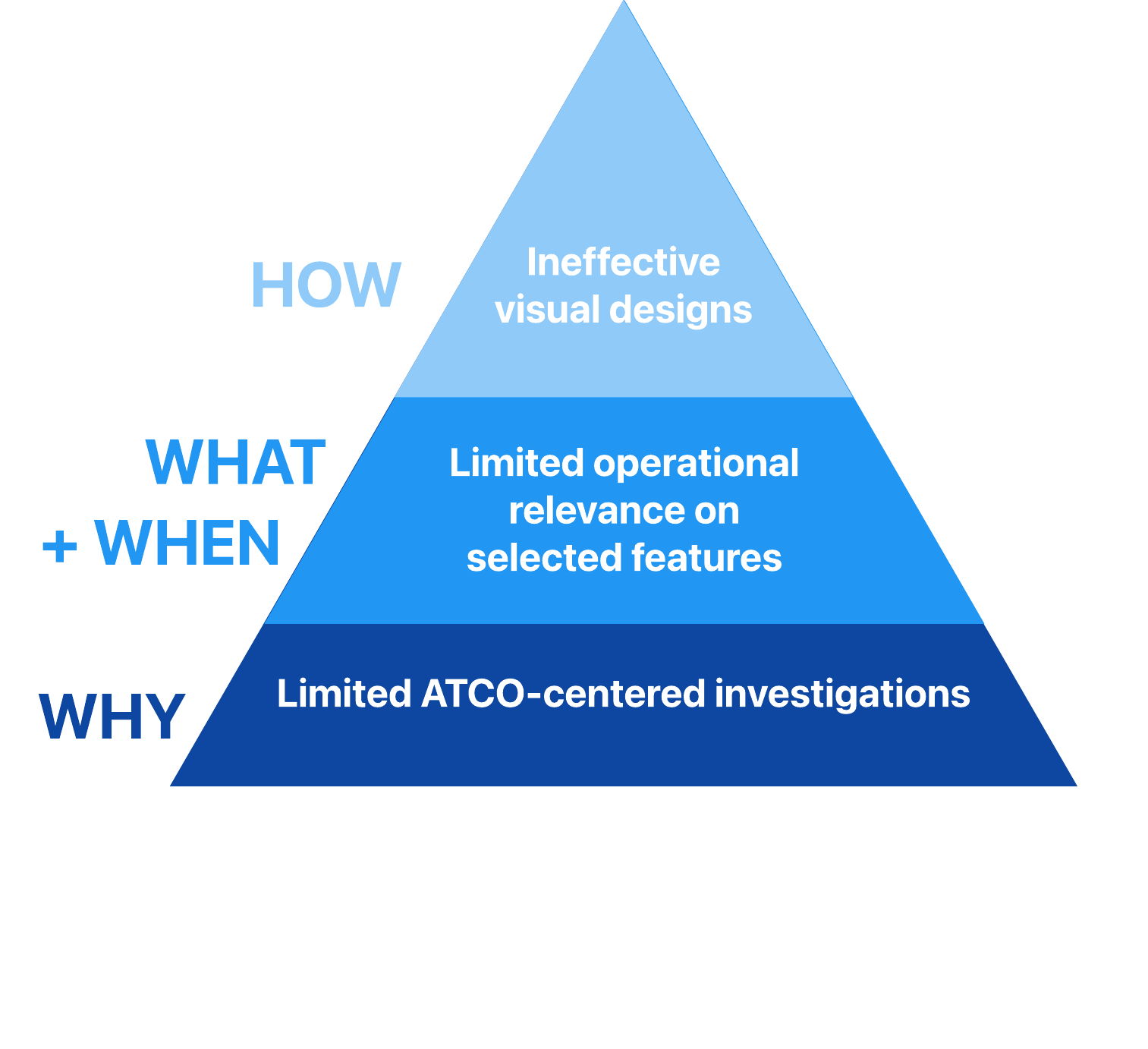}
    \caption{XAI research landscape in ATM. This study focuses on the foundational question of `why' ATCOs need explanations.}
    \label{fig:pyramid}
\end{figure}

\begin{figure*}[t]
    \centering
    \includegraphics[width=1.0\textwidth]{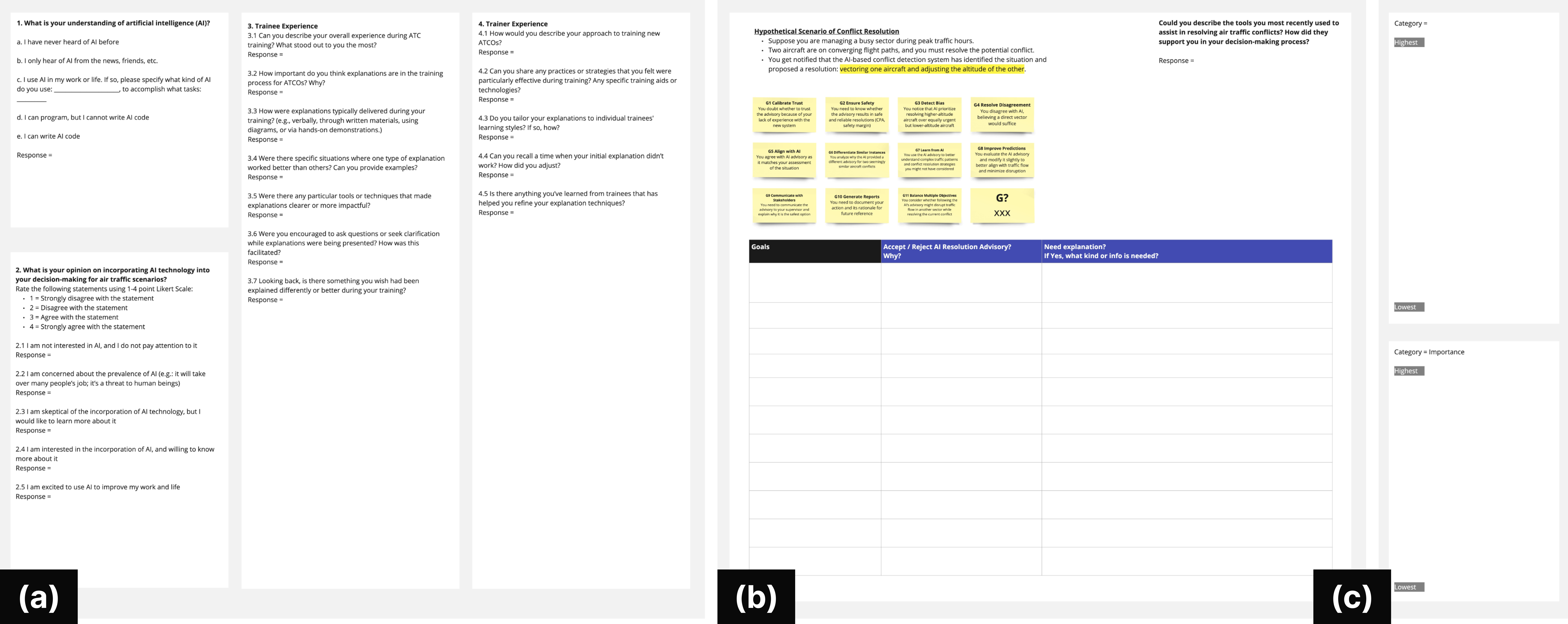}
    \caption{A virtual board used in the study, consisting of three activities: (a) semi-structured interviews, (b) goals exploration, and (c) goals ranking. Each goal is represented by a yellow card. Due to the limited text size in this figure, readers are recommended to view the higher resolution available at https://tinyurl.com/3mdwau4v}
    \label{fig:miro_study1}
\end{figure*}

At first glance, most existing studies emphasize the visual presentation of explanations~\cite{xie2021xai, valle2022TAPAS, hurter2022cdr, jmoona2023tot, pushparaj2023xai}.
Investigations by Hurter et al.~\cite{hurter2022cdr} and Pushparaj et al.~\cite{pushparaj2023xai} revealed that ATCOs preferred baseline `black box' conditions where no explanations were provided.
While surprising, this result may represent only the surface of a deeper issue.
For instance, insights from Jmoona et al.~\cite{jmoona2023tot} noted a critical challenge: operational irrelevance of the features selected to construct explanations.
No matter how advanced or clear the visualizations are, user comprehension remains limited if the `what' of explanations fails to align with practical considerations.
In addition, Valle et al.~\cite{valle2022TAPAS} considered not only the `how,' but also the `when' of explanations, emphasizing that operational context influences the cognitive resources available for ATCOs to process them.
Building on these insights, we take a step further by reframing the discussion to focus on ATCOs' specific user goals in various operational contexts, addressing the fundamental question of `why do ATCOs need explanations in the first place?'
Only by answering this can we determine which elements of explanations effectively align with their goals.
Figure~\ref{fig:pyramid} summarizes the limitations identified in relevant previous research.

Therefore, the following research questions (RQs) were formulated using a bottom-up approach:
\begin{itemize}
    \item \textbf{RQ1: Why} do ATCOs need explanations?
    \item \textbf{RQ2: When} are explanations desirable or appropriate?
    \item \textbf{RQ3: What} constitutes an effective explanation?
    \item \textbf{RQ4: How} should explanations be presented to ATCOs for easy and accurate comprehension?
\end{itemize}

As a first step, the focus is placed on addressing the fundamental question of \textbf{why} ATCOs need explanations, which inherently sheds light on the \textbf{when}, \textbf{what}, and \textbf{how} of explanations for future deeper investigations.

\section{Method}

    \subsection{Participants}
    We recruited online a total of eight licensed ATCOs (two female, six male), aged 31 to 48 (\mean{34.5}, \sd{5.6}) from four nations across three continents.
    Two held Area control ratings (both radar and non-radar), three were rated for Approach control, two held dual ratings for Approach and Aerodrome control, and one was rated with all three control types (Area, Approach, Aerodrome).
    Mean ATC experience was 7.8 years.
    Three of the eight were ATC instructors.

    \subsection{Procedure}
    We conducted the study via a video conferencing platform, beginning with ATCOs familiarizing themselves with the Miro interface using their personal desktops or laptops.
    Miro (\url{https://miro.com/}) is a virtual collaborative platform enabling participants to digitally sketch their ideas while simultaneously verbalizing their thoughts.
    ATCOs were instructed to share their screen, and the sessions were both audio- and screen-recorded for subsequent qualitative analysis.
    The study had three main activities (see Fig.~\ref{fig:miro_study1}), designed to elicit both qualitative and quantitative responses from ATCOs, regarding their goals for needing explanations and the specific phases of their operation where such needs arise.
    Each ATCO session lasted 2 to 3.5 hours, with ample breaks recommended to prevent fatigue.

        \subsubsection{Interviews}
        We designed a variety of questions to guide our semi-structured interviews.
        Initially, we used multiple-choice question and Likert-scale ratings to assess ATCOs' initial understanding and acceptance of AI.
        We then delved deeper into their prior experience as trainees and where applicable, as trainers (or instructors).
        For trainee experiences, we asked open-ended questions to understand the role of explanations in their learning journey towards becoming rated controllers. 
        For example: How were explanations delivered? Were there specific situations or tools that helped them to interpret explanations more effectively?
        For trainer experiences, we also used open-ended questions, aiming to uncover how their perspectives on explanation processes and outcomes evolved, when transitioning from receiving explanations as trainees to generating them as trainers.
        Key questions included: What strategies or tools did they find effective for delivering explanations to trainees? How did they adapt their explanations to trainees' individual learning styles, or when initial explanation were ineffective?
        For more details, please refer to the Miro link in Fig.~\ref{fig:miro_study1}.

        \subsubsection{Goals Exploration}
        Each ATCO began by describing the most recent tool they used to resolve air traffic conflicts and its role in their decision-making.
        We introduced a hypothetical scenario in which they managed a busy sector during peak traffic, assuming an AI system detected a conflict and proposed a resolution, specifically: vectoring one aircraft and adjusting another's altitude.
        This scenario was reiterated before introducing each of the 11 explanation goals, adapted from~\cite{jin2021practice} which focused on non-ATM use cases, such as estimating house prices, predicting diabetes risk, buying a self-driving car, and preparing for a bird-knowledge exam.
        Unlike \cite{jin2021practice}, we retained all goals to validate their applicability in ATM through direct ATCO feedback.
        The adapted goals covered diverse ATM contexts, spanning varying levels of time-criticality, performance variability, and complexity.
        For each goal, we explained its meaning and asked ATCOs 1) whether they would accept AI as decision support and why, 2) whether they would need explanations, and 3) if so, what specific explanations they would need.

        The following goals were presented in a different random order to each participant:\\
        \keys{G1} \textbf{Calibrate Trust:} You doubt whether to trust the advisory because of your lack of experience with the new system.\\
        \keys{G2} \textbf{Ensure Safety:} You need to know whether the advisory results in safe and reliable resolutions (CPA, safety margin).\\
        \keys{G3} \textbf{Detect Bias:} You notice AI prioritize resolving higher-altitude aircraft over equally urgent lower-altitude aircraft.\\
        \keys{G4} \textbf{Resolve Disagreement:} You disagree with AI, believing a direct vector would suffice.\\
        \keys{G5} \textbf{Align with AI:} You agree with AI advisory as it matches your assessment of the situation.\\
        \keys{G6} \textbf{Differentiate Similar Instances:} You analyze why AI provided different advisory for two seemingly similar conflicts.\\
        \keys{G7} \textbf{Learn from AI:} You use the AI advisory to better understand complex traffic patterns and conflict resolution strategies you might not have considered.\\
        \keys{G8} \textbf{Improve Predictions:} You evaluate the AI advisory and modify it slightly to better align with traffic flow and minimize disruption.\\
        \keys{G9} \textbf{Communicate with Stakeholders:} You need to communicate to supervisor and explain why it is the safest option.\\
        \keys{G10} \textbf{Generate Reports:} You need to document your action and its rationale for future reference.\\
        \keys{G11} \textbf{Balance Multiple Objectives:} You consider whether following the AI advisory might disrupt traffic flow in another sector while resolving the current conflict.
        
        \subsubsection{Goals ranking}
        After all the goals were added to the table (see Fig.~\ref{fig:miro_study1}b), ATCOs were asked to sort them in one open and one closed category.
        For the open category, ATCOs could create and define their own categories based on their interpretation of the goals.
        For the closed category, or if they struggled with the open category, we asked them to prioritize the goals based on their importance as an ATCO, ranking them from most to least critical.
        Note that we informed ATCOs that the order did not need to be linear: multiple goals could share the same rank if they perceived them to be of similar importance.
        Regardless of the category, the sorting exercise allowed ATCOs to express the relative difference between each goal.
        They were also encouraged to think aloud during the process, create new goals on a blank card, and include the new card(s) in the sorting exercise.

    \section{Results}
    
        \subsection{Interview: \textbf{Perception of AI}}
\begin{figure}[t]
            \centering
            \includegraphics[width=\columnwidth]{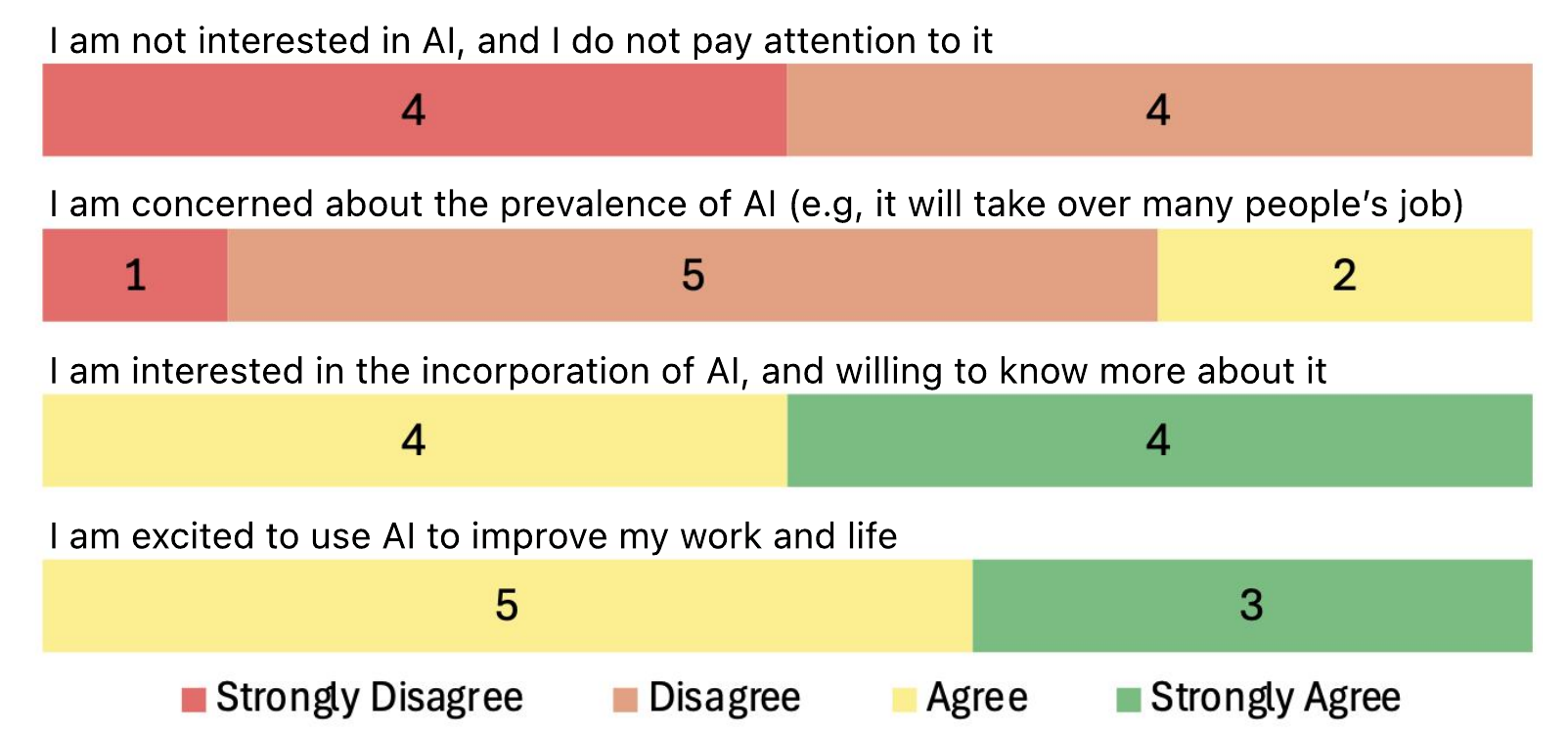}
            \caption{Distribution of opinions on incorporating AI technology into decision-making for air traffic scenarios}
            \label{fig:results_study1_profile}
        \end{figure}

        When asked about their understanding of AI, two ATCOs reported having only heard about it through news or friends.
        The other six had used AI tools like ChatGPT, Copilot, or Meta AI for tasks like proofreading emails, drafting job applications, creating travel itineraries, generating images, or clarifying technical ATM concepts.
        Only two ATCOs had used ChatGPT to write VBA code or to learn Python and SQL, and none had experience writing AI code.
        When asked about their acceptance of AI across the four statements (see Fig.~\ref{fig:results_study1_profile}, the majority disagreed with negative sentiments (reflected in the first two statements) and agreed with the positive sentiments (reflected in the last two statements).
        However, two ATCOs expressed concerns about the increasing prevalence of AI.
        
        \subsection{Interview: \textbf{Explanations during training}}
        All ATCOs unanimously reported that explanations are very important during their training for several reasons.
        Detailed explanations help \pquote{ease the transition for trainees}{P1}, particularly as they \pquote{face a big hurdle to climb to understand how to overcome the challenges}{P1}, despite the best efforts of seasoned instructors.
        One ATCO likened ATC to \pquote{a form of art}{P8}, arguing that explanations enable trainees to ``appreciate'' the nuance of decision-making.
        However, ATCOs also noted variability in explanation approaches, observing that \pquote{each trainer explains the same scenario differently}{P2,5,8} with some relying on instinct and personal experience, while others used quantitative facts (e.g., numbers, calculations) for clarity~{\small[P2]}.
        Trainers also struggled to tailor explanations, as one remarked, \pquote{once you’ve passed a certain hurdle, it’s very difficult to go back and explain it from a beginner’s perspective}{P1}, likening this challenge to explaining automatic skills like walking~{\small[P1]} or driving~{\small[P8]}.

        Explanations were delivered through various methods.
        In theoretical training, knowledge was primarily delivered through written materials.
        In simulator sessions, however, the approach becomes more dynamic.
        Trainers may \pquote{pause the session to explain}{P1} when the scenario becomes too complex and rely on real-time feedback by ``watching the trainee’s reactions'' while engaging in ``a lot of pointing and talking''.
        Trainees could also \pquote{pause and ask for immediate clarifications}{P8}.
        Some sessions were even recorded, allowing trainees to \pquote{hear their own voice and reaction}{P1} later for reflective learning.
        In on-the-job training (OJT), trainees were paired with a rated ATCO to manage live traffic together.
        Trainers documented explanations by \pquote{writing down on a form, sometimes drawing the scenario … then verbally explaining their observations}{P1}, serving as an audit tool to track progress.
        Flow charts and radar screens supplemented these methods, illustrating \pquote{what went wrong, and what could be improved}{P3}, while also acting as a \pquote{memory aid}{P4} and as a \pquote{visual representation of the verbal explanation}{P5}.
        Verbal explanations were primarily provided either on-the-spot or post-session in both simulation and OJT phases.

        The effectiveness of these methods depended on trainee learning styles and scenario complexity.
        For instance, some trainees rely heavily on the ``calculation method'' early on due to their lack of experience, while others benefited more from learning through experience, as they struggled with quick mental calculations~{\small[P2]}.
        In rapidly evolving scenarios, verbal explanations combined with visual cues—such as pointing to the radar screen—are preferred because \pquote{written instructions would take too long in a dynamic environment}{P1}.
        Additionally, the timing of explanations is critical.
        On-the-spot explanations benefited trainees with sharper short-term memory, whereas others preferred to have time to \pquote{internalize the comments}{P4} during post-session reviews when they are less overwhelmed.

        Generally, the training environment seems to encourage questions and clarification.
        However, there is an undercurrent of hesitation for some trainees who may fear being judged or making a poor impression, with one noting that \pquote{if you have a question they consider foolish, you'll still be judged}{P5}.
        This highlights a tension in the training culture, where the desire to ``make a good impression'' can sometimes discourage open inquiry for explanations.
        
        \begin{figure}[t]
            \centering
            \includegraphics[width=\columnwidth]{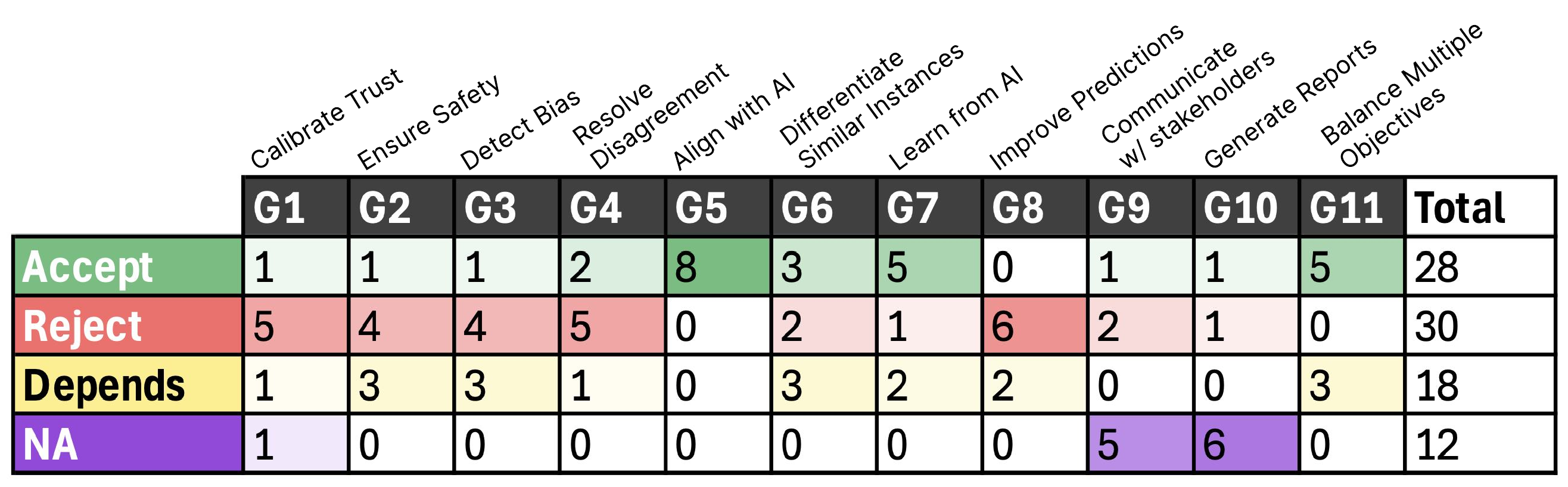}
            \caption{Response counts for each goal when a resolution advisory is presented.}
            \label{fig:results_study1_adv}
        \end{figure}
        
        In summary, the interview revealed that explanations are not merely supplemental but central to ATCO training.
        They bridge theory and practice, refine decision-making techniques, and ensure trainees not only learn the ``how'' but also understand the ``why'' behind critical decisions.
        The varied delivery methods—verbal, written, visual, and recorded sessions—reflect a multimodal effort to accommodate diverse learning styles and situational demands, ultimately to produce ATCOs who are both proficient and adaptable.
        
        \subsection{Goals Exploration: \textbf{Advisory Assessment}}

        We asked ATCOs whether they would accept AI as decision support and why, based on each of the 11 explanation goals.
        Since majority of the details do not centre around the need of explanations, we have made them available as supplementary content via https://tinyurl.com/pk6yxvcx.
        Here, we summarize the high-level insights as depicted by Fig.~\ref{fig:results_study1_adv}.

        All ATCOs unanimously accepted AI recommendations when their assessments aligned~\keys{G5} with the AI's.
        In cases of disagreement~\keys{G4}, rejections were more common than acceptance, driven by trust in personal skills, concerns over transmission length, or perception of sub-optimality, though some accepted if there is an overlap in assessment of the situation.
        For balancing multiple objectives~\keys{G11}, more ATCOs accepted the advisory due to the perceived capability for AI to manage complexity more efficiently than oneself.
        However, three said their decision depended on `\pquote{whether the AI’s explanation makes sense}{P5}, or if they \pquote{could think of a better solution than the advisory}{P6}.
        When the goal is to learn from AI~\keys{G7}, five ATCOs accepted the advisory to assess outcomes and leverage AI's optimization, two hesitated due to potential judgment overlap or time constraints, and one rejected it as distracting from their own plan.
        For \keys{G1},~\keys{G2}, and~\keys{G3}, which capture ATCOs' uncertainties about AI, most rejected the advisory—citing reliance on personal judgment and safety concerns—while a few conditionally accepted it to fill experience gaps or balance efficiency with safety.
        For \keys{G6}, responses were mixed: some accepted the context-dependent variations in AI outputs, others remained conditional pending further explanation, and a few rejected them in favor of consistent personal judgment.
        Most ATCOs found \keys{G9} and \keys{G10} not applicable in influencing advisory acceptance or rejection, citing the lack of live traffic, accountability concerns, and preference for independent decision-making, though a few accepted the advisory as a contingency for safety or reporting.

        \subsection{Goals Exploration: \textbf{Explanation Expectation}}

        \begin{figure}[t]
            \centering
            \includegraphics[width=\columnwidth]{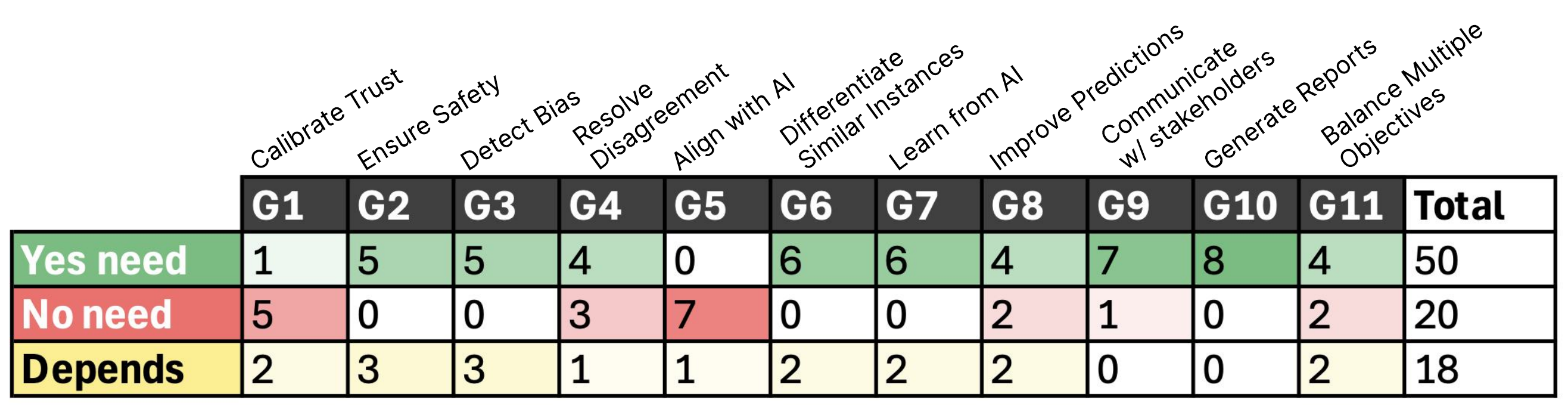}
            \caption{Response counts for each goal when asked if they need explanation.}
            \label{fig:results_study1_exp}
        \end{figure}

        All ATCOs needed explanations for both timely communication with stakeholders~\keys{G9} and post-event documentation~\keys{G10}, except for one in \keys{G9}, who believed that \pquote{while AI could solve immediate problems, its [advisory] might create new issues within 2–3 minutes}{P7}.
        Explanations were seen as vital for multiple reasons.
        Firstly, they help ATCOs comprehend \pquote{how a scenario evolved into a conflict, what was done, and what could have been done better}{P1}.
        Some ATCOs needed contextual explanations to seek clarity on decision-making parameters: \pquote{to better understand the considerations involved and why they are being applied}{P2}, reflecting a broader theme of transparency. 
        One even wondered if \pquote{AI can clearly articulate [their] thoughts in real-time}{P8} to support their discussion with supervisors, suggesting potential value of AI in enhancing human communication.
        
        Secondly, ATCOs stressed that AI explanations should complement, not overshadow, their judgment.
        For instance, they emphasized retaining agency, stating \pquote{I want my point of view in the report, not AI fully controlling the writing}{P3,4}.
        Some also acknowledged AI's potential to craft protective narratives, noting: \pquote{AI could write it in a way that protects me}{P1,4}.
        Overall, AI's explainability was viewed as a tool to \pquote{explain why [their] choice is the safer option}{P5}.
        One ATCO suggested AI could \pquote{reference specific regulations to support report writing}{P6}, further substantiating ATCOs' explanations.
        ATCOs also recognized AI's potential to address cognitive challenges, with one noting, \pquote{during moments of shock, it is difficult to articulate thoughts clearly}{P8}, suggesting that AI could \pquote{offer an objective perspective}{P8} of the situation.
        They proposed using AI to streamline reporting through visual playback and automated transcriptions, reducing administrative burdens: \pquote{AI could save significant time by automating manual tasks}{P8}.
        
        For \keys{G6} and \keys{G7}, six ATCOs expressed a need for explanations regarding AI decisions.
        However, two ATCOs {\small[P2,4]} specified that these explanations were more critical post-operation rather than during time-sensitive live operations.
        One ATCO explained, \pquote{there is no room for creativity [in real-time operation] as it can create unpredictability when working in a team}{P2}.
        They preferred executing resolutions that were easily accepted by colleagues, thereby minimizing the need for further explanation or justification.
        Another ATCO noted, \pquote{I might already have a clue about the AI's intention}{P4} from the advisory alone, reducing the need for additional explanations.

        In contrast, the remaining six ATCOs, who did not specify explanation timing, strongly desired to understand the rationale behind AI's varying decisions in \keys{G6}.
        One ATCO stated, \pquote{from my perspective, I don’t see any difference, but the AI sees something different, I want to know why}{P1}.
        Another added, \pquote{I want to know why the decision is different this time and what [factors were] taken into consideration}{P3}.
        This curiosity was driven by a need to align human and AI logic, as one ATCO explained \pquote{I want to see if [the AI] has the same understanding of the situation as I do}{P7}.
        Some ATCOs acknowledged human biases and sought explanations to \pquote{see if [they] missed anything, maybe [the instances] are not as similar as they seem}{P5}.
        In fact, an ATCO believed that it is \pquote{valuable for [AI to] anticipate future conflicts not yet visible to the ATCO}{P6}, and thus explanations would be needed to communicate the discrepancy.

        For \keys{G7}, the focus shifted to comparing \pquote{the difference between [one's] resolution and the AI advisory}{P7}, particularly in complex or unfamiliar scenarios.
        One ATCO described a case where \pquote{an aircraft needs to climb through three different aircraft approaching from different directions}{P1}, emphasizing the need to ``understand how the AI determines headings to resolve the situation''.
        Others echoed that without explanations, \pquote{you cannot learn the rationale}{P5} behind AI decisions.
        Another ATCO added, \pquote{the AI outcome may be due to variable A, or B, or a combination of B and C, so you won't know for sure, making it difficult to learn}{P6} without explanations.
        
        ATCOs expressed a strong need for explanations regarding how the system makes its decisions, particularly to address safety~\keys{G2} and bias~\keys{G3} concerns.
        For example, ATCOs emphasized the importance of \pquote{ensuring unsafe advisory does not result again}{P3} and \pquote{identifying and eliminating future biases}{P5}.
        In a similar vein, another ATCO noted the need to discern \pquote{whether the bias is really useful, or simply a result of system design}{P3}, while another added the need \pquote{to determine if it is a calculated risk or purely an unsafe recommendation or bias}{P5}.
        ATCOs also expected clear justifications when AI prioritizes one aircraft over another of equal urgency.
        P1 asserted that "AI should decide based on objective criteria, like time or speed, to break the tie", speculating that subtle factors, like a one-second time difference, might underlie these decisions.
        Others insisted that \pquote{there must be a good reason behind prioritization}{P2,3,7}.
        {\small P2} elaborated: ``Any conflicting aircraft should be treated equally, so if there is any to be prioritized, I need to know why'', and {\small P7} questioned, "is it because the higher [aircraft] has higher speed?''
        They argued that understanding safety hazard or bias in AI would help them \pquote{in extrapolating to other situations, detect biases in live operations, and inform trust in the system}{P5}.

        For resolving disagreement~\keys{G4}, improving predictions~\keys{G8}, and balancing objectives~\keys{G11}, half of the ATCOs needed explanations while the rest either did not need or were undecided.
        Explanations delivery boil down to subjective preferences.
        Some ATCOs found them unnecessary in resolved or time-constrained scenarios.
        An ATCO stated, \pquote{mostly if I reject [the advisory], I don't need explanation}{P2}, suggesting that explanations are less critical when a decision has been made.
        {\small P6} echoed this sentiment, explaining that ``since I am already improving it, I can see the problem,'' implying that real-time problem-solving often outweighs the need for detailed reasoning.
        ATCOs also prioritized swift decision making over understanding the AI’s rationale due to \pquote{the additional time needed to analyze reasoning}{P2}.

        Conversely, some ATCOs argued for the value of real-time explanations and the ability to control when explanations are presented.
        An ATCO believed that \pquote{explanations could completely change [their] decisions}{P3}, particularly when the AI identifies errors or gaps in the ATCO's initial assessment.
        This is because they believed that \pquote{AI can calculate things very quickly and accurately}{P5}, thus enabling them to determine which assessment—theirs or the AI's—is more optimal.
        In complex situations, explanations are \pquote{good to know what the AI is thinking when offering solutions with multiple objectives, because sometimes there might be too many objectives at once, and it could be something you missed}{P4}.
        Explanations were also valued in divergent situations, such as when \pquote{the AI offers a different advisory than what [they] initially thought}{P5}, or \pquote{whenever the AI can do better than [them]}{P3}.
        ATCOs sought clarity by asking, \pquote{What was its perspective? What did it see?}{P1}
        
        In situations where trust in AI is in doubt~\keys{G1}, five ATCOs found explanations unnecessary, citing that \pquote{distrust inherently introduces bias, making any AI explanation likely to be met with skepticism}{P4}.
        In fact, {\small P8} noted that ``the system might need more detailed explanation to convince [them] and build trust''.
        In contrast, an ATCO needed explanation \pquote{to be more comfortable and experienced with the system}{P3}, countering the lack of trust.
        Two ATCOs preferred on-demand explanations, suggesting they should be provided within a specific time window, such as \pquote{at least two to three minutes [before a conflict alert is triggered]}{P5}.

        Lastly, when ATCOs' assessments aligned with the AI's~\keys{G5}, all but one said they did not need explanations.
        One ATCO noted that they \pquote{had already formed an impression of the explanation}{P4} since \pquote{it already aligned with [their] goals}{P8}.
        The sole exception preferred on-demand explanation, stating that \pquote{some scenarios are not time-critical, and therefore there is time to view the explanations}{P2}—a useful option despite having their own reasoning.
        
        \subsection{Explanation Forms}
        Across the 11 goals, ATCOs expressed diverse preferences for how AI should deliver explanations, ranging from specific operational details, live interactions, and post-operation reviews.
        For specific operational details, they emphasized clear, concise data such as \pquote{tolerance thresholds and safety levels}{P1}, and insights into \pquote{why the decisions are made, and the thought process behind them}{P3}.
        One ATCO preferred textual/visual explanations over audio, stating, \pquote{I would rather see explanations because audio can be misheard or unclear}{P5}.

        During live operations, ATCOs stressed the importance of intuitive, accessible, and non-intrusive explanations.
        Several suggested interactive features, such as an \pquote{easy-to-click option to show the outcome of accepting a proposed advisory}{P6} or a \pquote{`Why?' button that triggers a textual explanation or visually highlight relevant traffic on the radar screen}{P2}.
        To avoid information overload, ATCOs recommended the ability to toggle explanations on and off, with concise messages like `Aircraft A heading xxx deconflicted with Aircraft B heading yyy,' noting that \pquote{simple keywords are sufficient, perfect English isn’t necessary}{P8}.

        For post-operation analysis, ATCOs preferred more detailed explanations.
        They suggested making explanations accessible \pquote{during radar playbacks when reviewing scenario details}{P6} and even \pquote{on a snapshot basis using specific timeframes}{P4}.
        One ATCO elaborated on the value of this approach: \pquote{it's good to have pictures and traffic context. If you just tell me to vector left, right, or center, I don't know what you're talking about. It's very hard to imagine}{P4}.
        Together, these insights indicate that ATCOs favor a flexible, context-sensitive delivery of explanations—one that supports immediate decision-making in live operations while also providing richer detail for retrospective analysis.
        
        \subsection{Goals Ranking}

        \begin{figure}[t]
            \centering
            \includegraphics[width=\columnwidth]{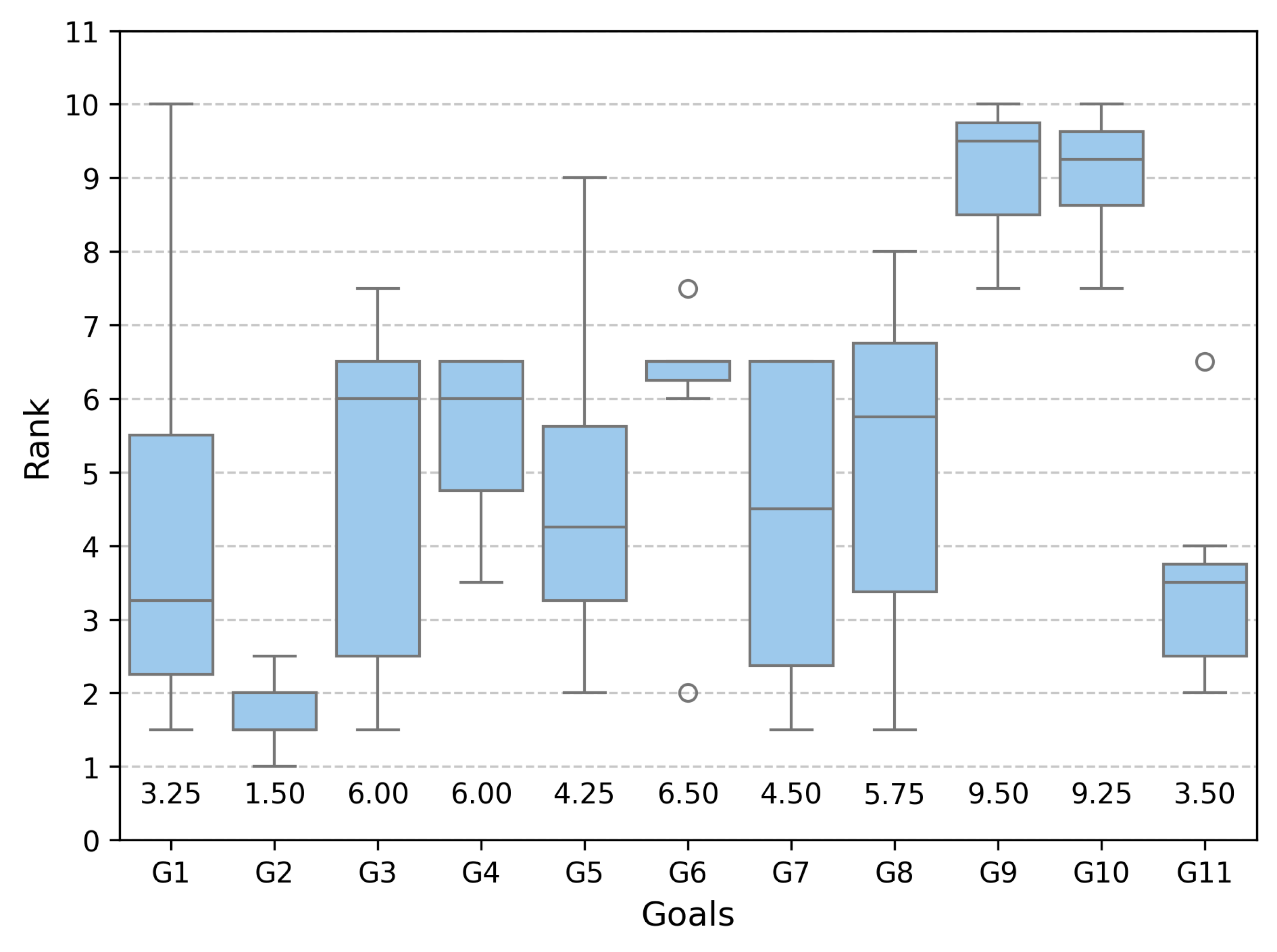}
            \caption{Rank distribution of goals based on perceived importance, with median values displayed below each box-plot. The lower the value, the more important it is perceived.}
            \label{fig:results_study1_rank}
        \end{figure}

        For quantitative analysis of ranked goals, we employed Friedman tests (non-parametric) which revealed no statistically significant difference (\chisquare{10}{8}{16.05}{=}{.098}) in rankings.
        This suggests ATCOs did not systematically prioritize the cards differently.
        While our small sample size (8 ATCOs) limits statistical power, descriptive statistics provide valuable insights.
        As shown in Fig.~\ref{fig:results_study1_rank}, \keys{G9} (\median{9.50}) and \keys{G10} (\median{9.25}) were perceived as the least important.
        ATCOs attributed this to the belief that such goals are \pquote{more suited for managers}{P2} and \pquote{do not directly impact the quality of actions in live operations}{P5}.
        Instead, these goals are better suitable for \pquote{reducing workload}{P2} and \pquote{time spent}{P8} on manual administrative tasks like communication and documentation.
        In contrast, most ATCOs consistently ranked \keys{G2} (\median{1.50}) as the most important, followed by \keys{G1} (\median{3.25}) or \keys{G11} (\median{3.50}) as the second most important.

\section{Discussion}
Synthesizing the results from the above interviews, goals exploration, and goals sorting, we discuss whether and why ATCOs need AI-generated explanations, summarize key takeaways for future investigations.

        \subsection{Explanations as Trust Builders}
        ATCOs naturally approach AI with a degree of skepticism, particularly in high-stakes situations where safety and human lives are at risk.
        This skepticism stems from a lack of familiarity with AI decision-making processes and concerns about the system's reliability~\keys{G2}\keys{G3}.
        Unlike the trainee-trainer relationship—where the trainer’s expertise is typically assumed and rarely questioned due to respect or authority (as mentioned in the interview)—trust in AI is not automatic.
        It must be earned through consistent demonstration of performance and transparency.
        To address this, explanations should clearly communicate the AI's decision making process, especially when discrepancies arise between ATCOs' judgments and AI recommendations~\keys{G4}\keys{G5}\keys{G6}.
        Without explanations, ATCOs are left to rely on their own `best guesses' about the system's logic which can be limited by their low trust~\keys{G1} and prior understanding of AI strengths and weaknesses.
        Trust in AI systems is built over time by fostering user confidence through evidence of reliable performance~\cite{jin2021practice}.
        Furthermore, Miller et al.~\cite{miller2019socialscience} emphasized the importance of contrastive explanations: humans seek explanations to understand why one event occurred over another.
        In the context of ATCO-AI interaction, this means explanations should not only clarify the AI's decision but also address why alternative options—particularly those considered by the ATCO—were not selected.
        This approach is especially critical during the initial phases of interaction or ATCO training, as it helps bridge the gap between human intuition and machine logic, ultimately building the foundation of trust in the system.
        
        \subsection{Explanations as Catalysts for ATCO-AI Collaboration}
        Once sufficient trust is established, the focus shifts from mere acceptance to optimizing collaborative decision-making.
        At this stage, ATCOs aim to work with AI as teammates, seeking explanations not only for understanding AI outputs but also refining their own judgments~\keys{G8} by considering alternative perspectives.
        As mentioned by {\small P5, P6}, and{\small P7}, some ATCOs need explanations to more decisively accept or reject advisories, particularly when human and AI judgments carry similar weight, or when the ATCO encounters an unfamiliar scenario.
        On the other hand, when AI advisories seem erroneous or unexpected, explanations aid in diagnosing potential system failures or inconsistencies, ensuring that ATCOs remain in control and never fall out of the decision-making loop.
        Central to this synergistic collaboration is recognizing and leveraging the complementary strengths of ATCOs and AI.
        AI systems excel at rapidly processing vast amounts of data, optimizing across multiple constraints~\keys{G11} and detecting subtle patterns that may be imperceptible to humans.
        Meanwhile, ATCOs contribute expertise grounded in experience, situational awareness, and stakeholder considerations~\keys{G9}—factors that remain difficult to fully encode into AI models.
        Given that explanation is inherently a social process, Miller et al.~\cite{miller2019socialscience} argue that humans tend to anthropomorphize AI and expect explanations that align with how they would explain human decision-making.
        Therefore, AI-generated explanations should not only be easily understood but should also be framed in ways that respect ATCOs' point of view and enhance their expertise~\keys{G10}, ensuring that automation remains a supportive tool rather than a replacement for human judgment.

        \subsection{Explanation as Tools for Coevolution}
        As ATCO-AI collaboration matures, the potential for over-trust and over-reliance on automated systems becomes a significant concern.
        Human decision-making is inherently susceptible to cognitive biases~\cite{wang2019theory}, such as confirmation bias, where individuals disproportionately favor information aligning with preexisting beliefs.
        In high-stress scenarios, the bias can cause ATCOs to default to AI advisories without sufficient scrutiny, jeopardizing safety.
        A static, one-size-fits-all explanation fail to capture the dynamic interplay between system capabilities and situational demands, leaving room for miscalibration of trust.~\cite{lee2004trust}.
        To address these challenges, we propose a shift toward adaptive automation coupled with context-aware explanations.
        Unlike static explanations, adaptive systems can adjust the level of automation based on factors such as ATCO’s situational awareness (e.g. SAGAT~\cite{endsley2017direct}), the complexity of the operational environment, and AI's confidence level.
        Such adjustments can mitigate the risk of both overtrust and distrust by encouraging ATCOs to reflect on their biases and recalibrate their trust to match the AI’s true capabilities or purposes~\cite{sanneman2022SA}.
        Grounded in cognitive psychology, these dynamic purpose-driven explanations and automation modes serve as cognitive tools that foster a continuous learning cycle—a coevolutionary process where both human and machine adapt and improve performance together.
        Over time, this balanced interplay ensures that ATCOs remain engaged decision-makers rather than passive recipients of automated advisories.\\
        
    \subsection{Operationalizing Effective Explanations in ATM}

    We highlight the following factors that had been raised by prior works in the context of our work, and what it implies for future work.

        \subsubsection{Timing (When)}
        There are three distinct phases during which XAI can be integrated: training, live operation, and post-operation.
        Our results align with Hurter et al.~\cite{hurter2022cdr}'s that XAI is more important during non-operation (i.e., before or after live operation).
        However, our results also reveal important nuances for each phase.
        During training, ATCOs typically exhibit lower operational understanding and trust in AI.
        Consequently, there is greater need for XAI that builds trust over time, tailored to each ATCO's level of experience and preferred learning style.
        Such XAI would address the challenge faced by human trainers, who often struggle to translate their experiential knowledge into accessible lessons for trainees.
        In live operations, on-demand XAI is necessary because not all scenarios are time-critical.
        ATCOs need explanations that enable them to make informed decisions by balancing with AI recommendations, or even integrating both.
        For post-operation activities, our goal exploration identified specific purposes like communicating with stakeholders~\keys{G9} and generating reports~\keys{G10}, where there is a surprisingly high demand for explanations.
        Future work could benefit from determining these timing options as scopes for deeper empirical investigations on XAI design.
        
        \subsubsection{Elements (What)}
        When constructing system-generated explanations, prior works relied on XAI methods (e.g. SHAP, LIME, and DALEX) to identify the top factors contributing to a system recommendation~\cite{jmoona2023tot, pushparaj2023xai}.
        In contrast, our findings show that a more operationally relevant approach could be to determine why ATCOs need an explanation.
        For example, is the need driven by concerns over safety buffers~\keys{G2}, or is it to differentiate seemingly similar conflicts~\keys{G6}?
        Future work could consider this by either interpreting these user goals through analysis of historical data, or by explicitly asking users with simple multiple-choice options.
        Such approach could help narrow down and tailor explanations to better meet their operational needs.

        \subsubsection{Format (How)}
        Current literature on XAI visualizations focused on static displays—either through direct rendering on the radar screen~\cite{hurter2022cdr}, or by presenting data charts on an external monitor~\cite{xie2021xai}.
        However, our user interviews suggested that a one-directional flow of information may not adequately support ATCOs' needs.
        Instead, we recommend the need for an interactive dialogue format, similar to their dynamic exchanges experienced during training sessions.
        This interactive approach would allow ATCOs to engage with the explanation actively—asking follow-up questions or requesting further details as needed—which could lead to a deeper understanding of AI decision-making processes.
        Moreover, given that six out of eight of the ATCO participants reported personal use of text-based natural language processing tools, such as ChatGPT, there is significant potential to explore conversational interfaces~\cite{abdulhak2024chatatc} for delivering explanations.

\section{Conclusion}
Previous research into XAI in ATM has predominantly focused on what explanations the system can generate, and how these explanations should be visualized based largely on researchers' intuition.
This research, on the other hand, reorients the focus towards the fundamental needs of the users: Do ATCOs need explanations, and if so, why?
By examining 11 operationally-relevant goals, the findings reveal that all ATCO participants needed explanations to document decisions and rationales for future reference or report generation.
However, explanations were deemed less necessary when there was an alignment between ATCO assessments and AI advisory.
From building trust, to catalyzing collaboration, and to supporting coevolution, the results indicate that explanations serve hybrid roles.
Importantly, this paper emphasized how explanations need to be dynamically adjusted to promote appropriate trust level, thereby mitigating distrust and overtrust.
Although these qualitative insights are valuable, the subjective nature of the research calls for further investigation with objective measures in future work.
In addition, while the primary focus was on conflict resolution scenarios, the insights may benefit future research in other operational contexts where long-term trust calibration is critical.
Overall, this investigation represents an initial step toward synthesizing prior research and advancing an ATCO-centered approach to XAI.

\section*{Acknowledgment}
This research is supported by the National Research Foundation, Singapore, and the Civil Aviation Authority of Singapore, under the Aviation Transformation Programme.
Any opinions, findings and conclusions or recommendations expressed in this material are those of the author(s) and do not reflect the views of National Research Foundation, Singapore and the Civil Aviation Authority of Singapore.
The authors thank Luke Neubronner for his valuable support in designing operationally-relevant scenarios, and also Duc-Thinh Pham and Thanh Danh Le for their insights and support.
This research has also been reviewed and approved by Nanyang Technological University’s (NTU) Institutional Review Board (IRB-2024-748).

\bibliographystyle{IEEEtran}
\bibliography{references.bib}

\end{document}